# Proper Time Foliations of Lorentz Manifolds

D. H. Delphenich†
Physics Department
University of the Ozarks
Clarksville, AR

*Abstract. Some standard definitions and results concerning foliations of dimension one and codimension one are introduced. A proper time foliation of Minkowski space in defined and contrasted with the foliation defined by the time coordinate. The extent to which a Lorentz structure on a manifold defines foliations, and the issues concerning the extension of the proper time foliation of Minkowski space to a Lorentz manifold are discussed, such as proper time sections of a geodesic flow.*

**0. Introduction.** Certainly much has been said about the *geometrical* consequences of Einstein's general theory of relativity [**1-4**]. This is because specifying a metric (or pseudo-metric) on a manifold implies the existence of a unique zero-torsion metric connection − the Levi-Civita connection − from which all of the rest of spacetime geometry − curvature, geodesics, conjugate points, etc. − follows by construction.

Less of a *physically* definitive nature has been said about how, or whether, spacetime *topology* differs from a Euclidean one ([1]). Certainly, the definition of a manifold weakens that assumption to being only locally true in general; however, the real issue is whether the process of raising time to the status of a dimension on a par with the other dimensions in the eyes of geometry is also true in the eyes of topology. A well-known result of general relativity [**5**] says that the mere existence of a maximal Cauchy hypersurface ([2]) Σ, as would be essential to expressing the metric of spacetime as a solution to an equation of time evolution, such as in the Hamiltonian formulation [**6**], implies that the topology of spacetime is simply the product $\Sigma \times \mathbb{R}$. This would tend to imply that the geometric unification of space and time is not topologically meaningful, which sounds suspiciously inconsistent with the very principle of (non-)simultaneity that supposedly spawned relativity to begin with.

Consequently, it may be worthwhile to suspend one's disbelief, however based in fact, about the necessity of generalizing the decomposition of spacetime into space and time, and examine the mathematical − and possibly physical − ramifications of starting with a more general requirement on spacetime than *global* hyperbolicity. The generalization we have in mind − that of foliations of manifolds − has already been subjected to considerable mathematical research activity ([3]) and a number of intriguing physical observations about the special case of foliations of Lorentz manifolds [**7-11**]. Nonetheless, for most physicists, the important issue is the Hamiltonian formulation of gravitation, in which case the theorem of Geroch seems to be yet another "no-go" theorem for the necessity of applying deeper topological methods, such as foliations and cobordism, to the problems of spacetime structure, since such generalizations would seem to apply only in a trivial case.

---

† ddelphen@ozarks.edu

[1]  Of course, the topological issues that relate to magnetic monopoles, wormholes, baby universes, and such, are *theoretically* important, but they nevertheless relate to things that have yet to show up in *experimental* practice, as some of the consequences of general relativity − bending of light rays, redshift, etc. − have.

[2]  Such a spacetime is defined to be *globally hyperbolic,* which is not to be confused with the fact that any Lorentz manifold has a *globally hyperbolic normal* pseudo-metric.

[3]  See Tondeur [**15**] for a *complete* (!) bibliography of *all 2500* books and papers on the subject up to the date of publication.

The basic result of this article is that there is a fundamental difference between the character of the foliation of Minkowski space that is defined by proper time level hypersurfaces and the foliations of Lorentz manifolds by the simultaneity hypersurfaces that one obtains by integrating a (presumably integrable) spacelike sub-bundle that is complementary to the line bundle that defines the Lorentz structure: they are locally concordant, but not always globally so. Indeed, the foliation of a Lorentz manifold that is obtained from the complementary spacelike sub-bundle has more in common with the foliation of Minkowski space by its time coordinate, which is a non-relativistic way of considering simultaneity.

Two classes of problems in which proper-time simultaneity play a significant role are the issue of whether the flows of extended matter admit sections – spacelike submanifolds transversal to the flow that "synchronize" the individual parameterizations of the curves – and the formulation of the Cauchy problem for hyperbolic systems on spacetime. The latter problem is not only concerned with the time evolution of localized waves of various sorts, such as electromagnetic and gravitational, but the cosmological problem of whether the spacetime pseudo-metric itself can be obtained from the time evolution of a metric that is defined on a spacelike three-dimensional submanifold of spacetime. As pointed out above, it is precisely this attempt to make the spacetime manifold a Cauchy development of a space manifold $\Sigma$ that implies that the resulting topology is "cylindrical," i.e., $\Sigma \times \mathbb{R}$.

## 1. Foliations [12-17].
A *foliation* $\mathscr{F}$ of a manifold M is a partitioning of M into submanifolds of dimension $n-k$, called *leaves,* i.e.,

$$M = \cup_\alpha \Lambda_\alpha, \qquad \Lambda_\alpha \cap \Lambda_\beta = \varnothing \qquad \text{if } \alpha \neq \beta,$$

in such a way that instead of an atlas of local diffeomorphisms onto $\mathbb{R}^n$, we have an atlas $\{U_\gamma, x\}$ of local submersions onto $\mathbb{R}^k$:

$$x: U \to \mathbb{R}^k, \qquad p \in M \mapsto (x_1, \ldots, x_k).$$

On the overlap of two foliation charts $\{U, x\}$ and $\{V, y\}$ there is a diffeomorphism $\phi: \mathbb{R}^k \to \mathbb{R}^k$ that makes $y = \phi \circ x$. This has the effect of taking subsets of leaves to subsets of leaves. One calls $k$ the *codimension* of the foliation, as opposed to *n-k,* which is its *dimension.*

In general, the morphisms that are associated with foliations should have the property that they take leaves to leaves. Clearly, this implies the foliations should have the same codimensions. Hence, if $(M_1, \mathscr{F}_1)$ and $(M_2, \mathscr{F}_2)$ are foliated manifolds with the same codimension then a morphism will be a differentiable map $f: M_1 \to M_2$ such that if $\Lambda \in \mathscr{F}_1$ then $f(\Lambda) \in \mathscr{F}_2$.

This implies that we have a reduction of the structure group for T(M) from GL($n$) to GL($n$-$k$)×GL($k$), or equivalently, a reduction of the bundle of linear frames GL(M) to the bundle GL($n$-$k$,$k$)(M) of frames whose first $n$-$k$ elements are tangent to the leaves and whose last $k$ elements are normal to it.

One aspect of this reduction is the fact that we can also define two sub-bundles of T(M): the rank $n$−$k$ vector bundle $\tau$(M) of tangent spaces to the leaves and $\nu$(M) = T(M)/$\tau$(M), the normal bundle to the leaves. However, going in the opposite direction, i.e., starting with the sub-bundle $\tau$(M) and looking for a partitioning of M is not straightforward at all. Indeed, that process is closely related to the problem of integrating differential systems, or, more generally, of integrating G-structures on M, where a G-structure is a reduction of GL(M), the bundle of linear frames ([4]), to a sub-bundle G(M) whose structure group G is a subgroup of GL($n$). G-structures subsume metric and pseudo-metric structures, orientations, differential systems, symplectic structures, and essentially any geometrically interesting structure on a manifold.

Before discussing the way that foliations relate to spacetime structure, let us look at some illuminating examples of foliations.

The simplest case of a foliation on a manifold is probably the case of a manifold of the form M = $\Lambda\times\mathbb{R}^k$ foliated by the projection onto $\mathbb{R}^k$, which has codimension $k$. Here, the leaves would be the submanifolds of the form $\Lambda\times(c_1, ..., c_k)$ and would therefore all be diffeomorphic. We will refer to foliations of the form $\Sigma\times\mathbb{R}$ as *cylindrical* and those of the form $\Sigma\times S^1$ as *circular,* which would physically correspond to either an infinitely extendible or periodic time parameter, respectively.

Since we have defined our notion of foliation in terms of an atlas of submersions, we naturally should mention that the prototype for a foliation is defined by the level surfaces to a submersion *f*: M → N. Locally, such a map looks like the projection of ($x_1$, ..., $x_{n-k}$, $x_{n-k+1}$, ..., $x_n$) onto its last $k$ components, where $k$ is the dimension of N. In particular, a non-singular smooth function on M defines a codimension-1 foliation of M; codimension-one foliations are one of the most important cases considered. An important example of a foliation by a submersion is that of a fibration, in which case, the leaves are the fibers. When a manifold is fibered over the real line or the circle, we are back to the case of a cylindrical, or circular foliation.

Foliations can be defined by almost-free ([5]) group actions, in which case the leaves are the orbits. However, this is not a necessary condition on the action; the natural action of SO(3) on $\mathbb{R}^3$−{0} is foliated by its orbits, even though the isotropy subgroup at any point is SO(2). This situation occurs often in the solution of Einstein's equations for cosmological models since a common assumption for such models would be various types of symmetry, such as spatial homogeneity and time invariance.

---

[4]  We use the notational convention that if G is a subgroup of GL($n$) then the corresponding reduction of GL(M) is notated G(M).

[5]  A group action G×M → M is called *almost free* if its isotropy group at any point is discrete, in which case the orbits all have the same dimension; however, they do not have to be diffeomorphic. If the action is *free* then the isotropy subgroups are all trivial and the leaves are all diffeomorphic to G.

Principal fiber bundles are foliated by the orbits of the action of the structure group, as well as by the level surfaces of the bundle projection.

A profoundly non-trivial example of foliation is defined by a non-singular dynamical system, which we will conceive in the form of a non-zero vector field X on a manifold. At every point $x \in M$, X spans a one-dimensional subspace of $T_x(M)$. The leaves of this one-dimensional foliation are the integral curves of X, where we are assuming that X is globally integrable, such as when M is compact. We can also consider this to be an example of an almost free action of the group $\mathbb{R}$ on M, where the isotropy subgroup is either $0$ or $\mathbb{Z}$, depending on whether the orbits (integral curves) are diffeomorphic to lines or circles.

In the hydrodynamical case, for which M is spacetime and the vector field **v** is presumed to represent the flow velocity of a matter continuum, the leaves of the dynamical system it spans are the *pathlines* of the flow. If M is cylindrical or circular: $M = \Sigma \times T$, with $T = \mathbb{R}$ or $S^1$, so the time direction can be singled out, then one can speak of a *steady* flow, for which $\mathbf{v}_\Sigma$, the spacelike part of **v**, is constant in time, and $\Sigma$ can also be foliated by the integral curves of $\mathbf{v}_\Sigma$, which are then called *streamlines*.

When spacetime is given a static non-zero electric (magnetic field, resp.) **E** (or **B**, resp.), the leaves of the dynamical system it defines are the field lines. The equipotential surfaces corresponding to non-zero static **E** or **B** fields define codimension-1 foliations, as do the velocity equipotentials for irrotational incompressible fluid flow when there are no stagnation points.

When a smooth function has no critical points, all of its level surfaces will be diffeomorphic. However, if a manifold is compact and boundaryless − such as a *compactification* of the spacetime manifold – then *any* continuous function must have a maximum and a minimum, hence, if it is differentiable, it must have critical points. An interesting class of critical points is the class of *nondegenerate* critical points, which are the ones for which the Hessian of the function has all non-zero eigenvalues at that point; by the inverse function theorem, such points are necessarily isolated. The fundamental theorem of Morse theory says that when a function $f$ on a manifold crosses a critical value $a$ the topology of $f^{-1}(x < a)$ changes by the attachment of a $k$-cell (or $k$-handle), where $k$ is the number of negative eigenvalues in the Hessian of $f$ at the critical point $f^{-1}(a)$.

Since the case of smooth functions on M with nondegenerate critical points, i.e., *Morse functions,* is so intimately related to the topology of M, it is reasonable to wonder what one obtains when one generalizes to foliations for which one drops the requirement that the charts be submersions. The answer is: *singular foliations,* or *Haefliger structures* [**18**]. According to Sardanashvily [**9-11**], such a foliation of spacetime would account for the singularity structure of its geodesics (conjugate points, focal points, caustics, etc.) as projective singularities of the geodesic flow on the cotangent bundle. In this way of looking at things, a geodesic field is described by a Lagrangian submanifold of the cotangent bundle, so deformations of that submanifold that do not preserve the fibration will produce singularities of the geodesic flow upon projection. In the electromagnetic example above, allowing **E** and **B** to have zeroes puts us into the context of Haefliger structures, as with fluid flow involving stagnation points; in any event, one is considering dynamical systems with fixed points.

A Haefliger structure amounts to defining an atlas of charts on a manifold M that consist of smooth maps of open sets of M onto $\mathbb{R}^k$, where $k$ is the codimension of the foliation. A second difference between this sort of atlas and the non-singular one is that the transition maps from one chart to another are only expected to be *local* diffeomorphisms of $\mathbb{R}^k$, not diffeomorphisms. One obtains leaves with singularities, and corresponding normal bundles with singularities. In the case of dimension one, a vector field with zeros will define a corresponding Haefliger structure. In the case of codimension one, a Haefliger structure makes the manifold look like it is locally foliated by level surfaces of Morse functions. One other aspect of Haefliger structures, which is of considerable use in the theory of characteristic classes of foliations, is the way singular foliations pull back under smooth maps as opposed to non-singular foliations, for which one must have that the map is tranversal to the leaves in order for the leaves to pull back to other leaves.

When one is given the sub-bundle $\tau(M)$ of $T(M)$, determining whether it is tangent to the leaves of a foliation of M is question of the integrability of $\tau(M)$, when regarded as a differential system on M. The main tool for considering the problem of integrability is Frobenius's theorem, which admits various formulations.

Suppose we notate the Lie algebra of vector fields on M by $\mathfrak{X}(M)$. In its tangent bundle formulation, Frobenius says that a *differential system* $\tau(M)$ on M, i.e., a sub-bundle of $T(M)$, is integrable iff the linear subspace $\mathscr{A}(M)$ of $\mathfrak{X}(M)$ that consists of vector fields on M whose values are in the fibers of $\tau(M)$ is a Lie sub-algebra of $\mathfrak{X}(M)$. In other words, iff:

$$X,\ Y \in \mathscr{A}(M) \Rightarrow [X,\ Y] \in \mathscr{A}(M).$$

This property is generally referred to by saying that $\tau(M)$ is *involutive* in that case.

There are various dual formulations of Frobenius in terms of exterior differential forms. The first one begins by representing the fibers of the sub-bundle $\tau(M)$ as the annihilating subspaces for an *exterior differential* system on M, namely, a set of $p$-forms $\theta^1, ..., \theta^k$, where $k$ is the codimension of $\tau(M)$, i.e., the dimension of its normal bundle $\nu(N)$. (This set is not unique, however. More generally, one usually regards the ideal $I\{\theta^1, ..., \theta^k\}$ in $\Lambda^*(M)$ that is generated by the set $\{\theta^1, ..., \theta^k\}$ as the exterior differential system.) One easily sees that the involutivity of vector fields in $\mathscr{A}(M)$ is equivalent to the statement that for each $i = 1, ..., k$:

$$d\theta^i = \eta \wedge \theta^i$$

for some 1-form $\eta$, which can also be written as:

$$d\theta^i = \tfrac{1}{2}\, c^i_{jk}\theta^j \wedge \theta^k,$$

where $c^i_{jk}$ are 0-forms on M ([6]). (One can also say that the integrability condition implies that the ideal $I\{\theta^1, ..., \theta^k\}$ is closed: $dI \subset I$.) This is, in turn equivalent to the statement that:

$$d\theta^i \wedge \theta^1 \wedge ... \wedge \theta^k = 0 \qquad \text{for all } i = 1, ..., k.$$

---

[6] If this looks suspiciously like the Maurer-Cartan equations for a Lie group G then that is because the Maurer-Cartan equations represent the integrability conditions for the "0-dimensional foliation" of G defined by integrating the tangent spaces, i.e., the Lie algebra of G, to obtain the points of G.

Note that for the normal bundle to be globally spanned by $k$ such forms, it must be trivializable. In this case one speaks of a *framed foliation.* Similarly, if the normal bundle is orientable then one speaks of an *oriented foliation.* Since every frame defines an orientation, being framed is a stronger requirement. However, in the case of codimension one, a non-zero vector field defines both a trivialization and an orientation on the normal bundle, so the two notions coincide for that case.

A corollary to Frobenius is that any one-dimensional foliation is integrable. This should not be confused with the previous observation that vector fields do not always admit global flows, since a global flow implies a specific *parameterization* of the integral curve. Neither does it imply that every one-dimensional foliation can be represented by a non-zero vector field, as we shall observe later in the context of time-orientability.

Let us return to the statement that when a codimension-one foliation is defined by a 1-form $\theta$ the subbundle $\tau(M)$ that is annihilated by $\theta$ [7] is integrable iff $d\theta^\wedge\theta = 0$ iff $d\theta = \eta^\wedge\theta$ for some $\eta \in \Lambda^1(M)$. It is not hard to show that $\eta^\wedge d\eta$ is closed since:

$$0 = d(d\theta) = d\eta^\wedge\theta - \eta^\wedge d\theta = d\eta^\wedge\theta - \eta^\wedge\eta^\wedge\theta = d\eta^\wedge\theta.$$

Hence $d\eta = \alpha^\wedge\theta$ for some 1-form $\alpha$, which implies that $d(\eta^\wedge d\eta) = d\eta^\wedge d\eta = 0$. This means that the 3-form $\eta^\wedge d\eta$ defines an element of $H^3(M, \mathbb{R})$, the 3-dimensional de Rham cohomology space of M, which is called the *Godbillon-Vey class* of the foliation, GV[$\mathscr{I}$] [**19,20**]. What is not obvious is that the class is indeed a "characteristic" of the foliation and not the manifold, and proving this was the reason that the result of Godbillon and Vey was so important.

To justify removing the quotes from the word "characteristic" implies that we have some way of classifying foliations of a manifold up to equivalence. In fact, there are a number of ways, of which we will mention three [8].

First, there is the possibility that two foliations $\mathscr{I}_0$: M = $\cup L_\alpha$ and $\mathscr{I}_1$: M = $\cup \Sigma_\beta$ of a manifold that have the same codimension may be *integrably homotopic*. This means that there is a foliation of M×[0, 1] of the same codimension such that:

  *i*) Its leaves intersect M×{$s$} transversally for all $s \in [0, 1]$.
  *ii*) When $s = 0$ or 1, the intersections are leaves of $\mathscr{I}_0$ or $\mathscr{I}_1$, respectively.

Second, there is the possibility that they are *concordant.* This a weaker form of integrable homotopy for which one substitutes the requirement that the leaves of the foliation on M×[0, 1] need to be transversal only for $s = 0$ and 1 for the first condition above. A consequence of this definition is that when two foliations are concordant, their normal bundles are homotopically equivalent. If two foliations of the same manifold are concordant then they will have the same Godbillon-Vey class.

---

[7] N.B. This implies that the normal bundle $\nu(M)$ would have to be *orientable,* a special case we shall return to in the context of Lorentz structures.

[8] For more details, cf. [**12**].

A third type of equivalence is (*foliated*) *cobordism,* in which two foliated manifolds $\mathcal{F}_0$ and $\mathcal{F}_1$ are equivalent if their disjoint union bounds a foliated manifold $\mathcal{H}$ of one higher dimension such that the foliation of $\mathcal{H}$ is transverse to the foliations of $\mathcal{F}_0$ and $\mathcal{F}_1$. (This is not the same as specifying that the boundary components are leaves of the foliation $\mathcal{H}$.) In the case of a closed orientable 3-dimensional boundary component we can integrate the Godbillon-Vey class to obtain its *Godbillon-Vey number.* This number is a foliated cobordism invariant. It is also interesting to observe that unlike other characteristic numbers, such as Chern and Pontrjagin, which take their values in the integers, the Godbillon-Vey number is not even rational. This implies a possible continuum of inequivalent codimension-one foliations on closed 3-manifolds. Indeed, Thurston [**21**] showed that such an infinitude does exist on $S^3$.

## 2. The proper time foliation of Minkowski space.
Although some of the foregoing constructions take on a trivial character when applied to the case of a vector space − since the tangent and cotangent bundles are trivial in that case − nevertheless, because the basic idea of differential analysis is to view nonlinear things as deformations of locally linear things, it is still worthwhile to briefly examine the vector space case, if only to build an intuition of expectations for the transition to nonlinear manifolds. In particular, the main issue is that of describing the global topological obstructions to the locally "trivial" concepts being globally trivial as well.

Just to set the record straight, in this article we will be using the convention that the Lorentz scalar product that defines a Minkowski space of dimension $n$ has one plus sign and $n-1$ minus signs ([9]), i.e., $n$-dimensional Minkowski space is $\mathbb{R}^n$ given the *normal hyperbolic* scalar product:

$$\eta(\mathbf{e}_i, \mathbf{e}_j) = \eta_{ij} = \mathrm{diag}(1, -1, \ldots, -1),$$

in which $\{\mathbf{e}_i, i = 0, \ldots, n-1\}$ is the canonical basis for $\mathbb{R}^n$.

The simplest way to define a codimension-one foliation on Minkowski space is to define the leaves as the affine subspaces that correspond to each $x_0 =$ const.; this is simply the foliation defined by projection onto the first coordinate. If the first coordinate represents a "time" coordinate, then this is essentially the Newtonian way of looking at simultaneity when light is presumed to travel at infinite speed. However, the Newtonian foliation has nothing to do with the Lorentz structure, which, of course, originates in the relativistic treatment of simultaneity in a world where light travels at finite speed. Hence, we might also want to look into foliations − or rather, Haefliger structures − that depend on the Lorentz structure, i.e., the scalar product, for their definition.

This scalar product defines a quadratic form by $Q(\mathbf{v}) = \eta(\mathbf{v}, \mathbf{v})$. The differential of this function is the linear functional on $\mathbb{R}^n$ defined by ([10]):

---

[9] The main reason for choosing this convention is to make the proper time parameter real for timelike vectors and imaginary for spacelike ones. In some cases, it does not change the qualitative nature of the results, though.

[10] We shall use the notation D for the differential of a map in order to avoid confusion with the *d,* the exterior derivative

$$DQ|_{\mathbf{x}}(\mathbf{v}) = 2\eta(\mathbf{x}, \mathbf{v}), \quad \text{i.e.,} \quad DQ = i_{\mathbf{x}}\eta,$$

which has a nondegenerate critical point at $\mathbf{x} = 0$ since the Hessian of Q is $D^2 Q = \eta$ everywhere. Hence, we can either delete the origin and foliate $\mathbb{R}^n - \{0\}$ by the level surfaces of the restricted submersion Q or deal with the singular foliation Q defined on all of $\mathbb{R}^n$; we shall pursue the latter course.

If one ignores the light cone, which is defined by the critical value of Q and is not even a manifold, then the leaves of the foliation defined by Q are its level surfaces (see Fig. 1). Topologically, there are two types of them (at least when $n>2$).

For $Q(\mathbf{v}) = \tau^2 > 0$, the level surfaces are *spacelike* hypersurfaces that are diffeomorphic to $\mathbb{R}^{n-1}$ by projection onto the $v_1 = 0$ subspace. Of course, we should not neglect to observe that for a given $\tau^2$ there are two such hypersurfaces, depending on the sign of $v_1$, i.e., which root of $\tau^2$ we choose. The positive values define the *future* and the negative ones, the *past*.

For $Q^2(\mathbf{v}) = -\tau^2 < 0$, the level surfaces are *timelike* hypersurfaces. We infer that $(v_1, ..., v_n)$ goes to the point $(\mathbf{r}, v_1^2 + \tau^2)$, where $\mathbf{r}$ is the projection of $(v_1, …, v_{n-1})$ onto the unit $n-2$-sphere. All values of $v_1$ give non-negative values for $v_1^2 + \tau^2$, so the image of a spacelike hypersurface is diffeomorphic to $S^{n-2} \times \mathbb{R}$. In this case $\tau$ is imaginary, which gives spacelike vectors a "non-physical" character, and the positive and negative imaginary roots of $-\tau^2$ both give the same hypersurface, rather than disjoint components.

Although the fundamental theorem of Morse theory [22] applies to only compact manifolds, we see that something similar is going on as we pass the critical value of $Q$, if we restrict $Q$ to a Euclidean ball about the origin. For $Q<0$, the inverse image, $Q^{-1}((-\infty, \tau^2])$, is homotopically equivalent to an $n-2$-sphere, since it is diffeomorphic to $(0, 1] \times (0,1) \times S^{n-2}$. As we pass the critical value $Q = 0$, which has Morse index $n-1$, we attach an $n-1$-cell (an $n-1$-*handle*, in Smale's terminology [23]) and find that $Q^{-1}((-\infty, \tau^2])$ becomes homotopically equivalent to that sphere with the center filled in, which is homotopically equivalent to a point. Graphically, we have the situation depicted in Fig. 2.

Had we chosen the opposite sign convention for the scalar product the sequence of spaces define by successive values of Q would be: a disjoint pair of $n$-cells (= two points, homotopically), the attachment of a 1-cell, a space homotopically equivalent to a point (see Fig. 3).

Now that we have two foliations (one of which is singular) for the same space, we should naturally ask if they are concordant. One the one hand, the fact that not all of their leaves are diffeomorphic sounds unpromising, but since $H^3(\mathbb{R}^n, \mathbb{R}) = 0$, they will both have the same Godbillon-Vey class. However, equality of the Godbillon-Vey classes is a necessary, but not sufficient, condition for concordance. It turns out that when we eliminate the light cone from $\mathbb{R}^n$ then the proper time

---

operator that acts on exterior differential forms.

foliation is actually concordant to two copies of the Newtonian foliation of $\mathbb{R}^n - \{x_0 = 0\}$. Since the transition from Lorentzian to Newtonian time is defined by the transition from $c = 1$ to $c = \infty$, it is illustrative to make the concordance explicit in terms of the passage to the Newtonian limit.

Note that the proper time foliation of either nappe of the open time cone is concordant to the foliation of the open upper half plane by the "time" coordinate $x_0$. First, we use spherical coordinates to express a point of a proper time hypersurface, so $(x_0, x_1, x_2, ..., x_{n-1})$ corresponds to $(x_0, r, \theta_1, ..., \theta_{n-2})$. Then, we pass to hyperbolic coordinates on the hypersurface by means of the transformation of the first two coordinates:

$$x_1 = \tau \cosh\alpha \qquad\qquad \tau = \pm\sqrt{x_0^2 - \left(\frac{r}{c}\right)^2}$$

$$r = \tau \sinh\alpha \qquad\qquad \alpha = \tanh^{-1}\left(\frac{r}{cx_0}\right).$$

We define a homotopy that "flattens out" the $\tau$-hypersurfaces into $x_0$-hypersurfaces as $c$ goes from 1 to $\infty$:

$$\text{F: } [0,1] \times \{\tau > 0\} \rightarrow \{x_1 > 0\}, \qquad (1/c, \tau, \alpha) \mapsto ((1-1/c)\tau\cosh\alpha + \tau/c, \alpha).$$

As $c \mapsto \infty$, $\tau \mapsto x_1$, and $\alpha$ takes on the interpretation of the Euclidian distance from the origin to the resulting point in the $x_0 = \text{const.}$ affine subspace. The hypersurfaces defined by the constant-$\tau$ maps:

$$(\alpha, 1/c) \mapsto \text{F}(1/c, \tau, \alpha)$$

define the leaves of a foliation of $[0,1] \times \{\tau > 0\}$; hence, we have our concordance of the open time cone with $\mathbb{R}^n - \{x_0 = 0\}$. In effect, as $c$ goes to infinity the light cone converges to the $x_0 = 0$ hyperplane.

To see how the timelike leaves outside the light cone are concordant to the Newtonian foliation of $\mathbb{R}^n - \{x_0 = 0\}$, note that in polar coordinates we have that the space of spacelike leaves of the proper time foliation looks like $\mathbb{R}^* \times \mathbb{R}^+ \times S^{n-2} \in \{\text{time line} - 0\}$, with coordinates $(\tau, \alpha, \theta, ...) \in (\tau, 0, 0, ...)$. By comparison, the space of timelike leaves looks like $\mathbb{R}^+ \times \mathbb{R} \times S^{n-2}$, with coordinates $(\tau, \alpha, \theta, ...)$. (Here, we have chosen the positive imaginary root of $\tau^2$ and redefined $\alpha$ so that now: $t = \tau\sinh\alpha$, $r = \tau\cosh\alpha$.) If we were to eliminate the $t = 0$ hyperplane then this would look the timelike case with $\tau$ and $\alpha$ reversed. Hence, if we eliminate the $t = 0$ hyperplane by a one-parameter rotation through $\pi/2$ in the $(\tau, \alpha)$ plane and add the time line back to the resulting leaves then we obtain the spacelike leaves of the proper time foliation, which we already show to be concordant to the Newtonian foliation of $\mathbb{R}^n - \{x_0 = 0\}$. In effect, we may think of the timelike and spacelike vectors of Minkowski space as playing the geometric role of real and imaginary numbers, respectively. Indeed, for the present choice of sign convention on $\eta$, the analogy is exact for the corresponding proper time coordinates.

The leaves of the proper time foliation can also be defined by the orbits of the action of $SO(1, n-1)$ on $\mathbb{R}^n$. The isotropy subgroups for the various leaves are not discrete, so the action is not almost-free, but for the timelike hypersurfaces, the isotropy subgroups are isomorphic, and similarly for the spacelike hypersurfaces. Hence, the open time-cone, as well as the spacelike complement of the closed time-cone, is foliated by orbits of the Lorentz group. In four dimensions, since time-like vectors are fixed by 3D rotations about themselves, the spacelike proper time orbits are diffeomorphic to $\mathbb{R}^3$. Since spacelike or lightlike vectors are fixed by a more elaborate combination of 3D rotations and boosts, the timelike or lightlike orbits are diffeomorphic to $S^2 \times \mathbb{R}$. The origin is fixed by everything so it is its own orbit. An important physical aspect of the proper time foliation is its Lorentz invariance, as opposed to the Newtonian foliation. This suggests the proper time foliation is more physically intrinsic than the Newtonian one.

The physical significance of the proper time leaves is that an observer placed at the origin of Minkowski space will have a local conception of proper time simultaneity defined by all spacetime points that can be reached by traveling at a given speed for a fixed amount of time. Here, we see a major distinction between the leaves: the spacelike leaves allow only speeds less than $c$, and travel into the future (or past) exclusively, whereas the timelike leaves demand speeds over $c$ and travel into either past or future. When the speed equals $c$, we are considering the points of spacetime that can only be reached by light rays. In this case, proper time is zero for all such points, so the "elapsed" time it takes to get a light ray from the observer to another point has to be measured by the $t$ coordinate.

If we exclude the light cone then we can find a one-dimensional foliation that is Lorentz-orthogonal to the level surfaces by taking the gradient of Q at any point $\mathbf{x}$ in $\mathbb{R}^n - \{\text{light cone}\}$, i.e., the Lorentz dual of the 1-form $DQ = 2i_{\mathbf{x}}\eta$, which is simply $2\mathbf{x}$. This vector field is clearly zero only at the origin, so everywhere else it generates a line field that is orthogonal to the tangent spaces to the proper time hypersurfaces. For the open time-cone, the leaves of this foliation are lines of the form $\tau\hat{\mathbf{r}}$ where $\hat{\mathbf{r}}$ is the unit vector in the direction of $\mathbf{x}$ and $\tau$ is the proper time parameter. These are the lines that an observer located at the origin at $\tau = 0$ would follow during unaccelerated motion, at least in terms of their own observations.

This gives us a convenient opportunity to introduce another way of defining the same foliation on $\mathbb{R}^n$ without introducing the metric directly: by choosing one of the lines L through its origin to represent the path that is followed by an observer "at rest;" by definition, it is timelike. Since the Lorentz group acts on $\mathbb{R}^n$ by way of $n \times n$ real matrices and takes timelike vectors to timelike vectors, one can obtain all of the other timelike lines by that action. In fact, one can also reconstruct a Lorentz metric from this choice of line: give $\mathbb{R}^n$ its canonical Euclidean metric $g_{ij} = \delta_{ij}$ and choose either unit vector $\mathbf{t}$ that generates L and call the dual 1-form $\theta = i_{\mathbf{t}}g$. By using the Euclidean metric, one can also speak of the $n$-1-dimensional subspace $\Sigma$ that is orthogonal to L, as well as the restriction $g_\Sigma$ of the Euclidean metric to it. One can then define a Lorentz metric by way of:

$$g = \theta \otimes \theta - g_\Sigma.$$

Of course, our choice of the canonical Euclidean metric is peculiar to $\mathbb{R}^n$, whereas in a more general vector space any other Riemannian metric would produce a Lorentz metric from a choice of L. Hence, there is nothing canonical (i.e., exists uniquely) about the association of a Lorentz metric with a choice of line through the origin. Conversely, as we just saw, a choice of Lorentz metric on a vector space defines only a family of timelike lines through the origin.

Notice that merely choosing a line L through the origin to represent the path of an observer at rest does not uniquely define the *proper-time parameterization* of this line either. Of course, one could use the Riemannian metric to parameterize L by arclength, but as we just pointed out, the choice of this metric was canonical only in $\mathbb{R}^n$.

Another point worth making is that, insofar as an "observer" is presumed to follow a single line or curve, such an observer is consequently presumed to have a pointlike nature. In order to accommodate the needs of extended observers, we need to go to the more general category of Lorentz manifolds, which we now proceed to do.

**3. Lorentz structures on manifolds.** The basic assumption about Lorentz manifolds is that they look like Minkowski space locally, i.e., in their tangent spaces. This means that we can define a Lorentz manifold M to be one that has a symmetric non-degenerate doubly-covariant tensor field $g$ such that for every point $x \in M$ there is some frame $\mathbf{e}_i$ in $T_x(M)$ such that $g(\mathbf{e}_i, \mathbf{e}_j) = \eta_{ij}$. Such a frame is called *orthonormal*. Since, by definition, any Lorentz transformation of the tangent space will take an orthonormal frame to another orthonormal frame, and the Lorentz group is not the identity group, the chosen frame is not unique, but belongs to an orbit of equivalent frames under the action of the Lorentz group. Indeed, there is a (non-canonical) one-to-one correspondence between orthonormal frames and elements of the Lorentz group, since the isotropy subgroup of this action on any frame is the identity. Restricting the linear frames at each point of M to the orthonormal frames at that point defines a reduction of the linear frame bundle GL(M) to the Lorentz frame bundle O(1,$n-1$)(M).

As we observed in Minkowski space, a Lorentz scalar product is equivalent to a choice of line through the origin. Hence, an equivalent way of defining a Lorentz structure on a manifold M is to define a *line field* L(M) on M ([11]). This amounts to a section of PT(M), the *projectivized tangent bundle* of M, whose fiber at a point $x \in M$ consists of all lines through the origin of $T_x(M)$.

In a sense, the line that one chooses defines the symmetry axis of the time cone at $x$; physically this choice represents a choice of rest frame at every point. More precisely, a *rest frame* at a point $x \in M$ is an *equivalence class* of Lorentz-orthonormal frames that all have a common timelike member; the equivalence relation is defined by the orbits of the action of SO(3) on SO(1,3)(M) at that point. This shows that a rest frame is also an SO(3)-structure on M. For a rest frame, the timelike leg represents the generator of the local time axis at that point; moreover, the proper-time parameterization of that axis agrees with the time coordinate parameterization for a class of chart. One would say that the rest frame is then *adapted* to the foliation of the coordinate chart by level surfaces of the time coordinate, which we identified as the Newtonian perspective on time and simultaneity. Clearly, every coordinate chart admits a rest frame.

---

[11] *cf.* Markus [**24**], Greub [**25**], and Steenrod [**26**]

In particular, the line field on M defines a metric that makes the line at every point timelike. To generalize the previous construction, note that if the line at $x$ is generated by a non-zero vector **t** and we choose a Riemannian metric $g_R$ on M (which always exists for paracompact M, but not canonically) so we can define the 1-form $\theta_x = i_t g_R$ and the orthogonal space $\Sigma$ at $x$ then we can define a Lorentz metric on $T_x(M)$ by:

$$g = \theta \otimes \theta - g_\Sigma.$$

Since we have tacitly assumed that **t** has unit norm, we should observe that this, in turn, assumes that M admits an everywhere non-zero vector field. If M is non-compact this is always the case, but if M is compact then the Euler-Poincaré characteristic must vanish. Furthermore, there is a difference between M admitting *some* non-zero vector field and M admitting a non-zero vector field *that generates the given line field*. We shall discuss this aspect of Lorentz structures later in the name of time orientability.

As before, the relationship between the choice of line and the Lorentz metric is not canonical. Similarly, the choice of line field itself is not canonical. Indeed, such sections fall into homotopy classes, the structure of which is a problem in obstruction theory that bears heavily upon the homotopy groups of the fibers of PT(M), which are isomorphic to the various $\pi_i(\mathbb{R}P^{n-1})$, but not canonically. Even for the case of a parallelizable manifold, i.e., a trivial PT(M) = M$\times \mathbb{R}P^{n-1}$, where the homotopy classes are simply [M, $\mathbb{R}P^{n-1}$], there is possibly more than one homotopy class. For example, when M=S$^3$ we have [M, $\mathbb{R}P^2$] = $\pi_3(\mathbb{R}P^2) = \pi_3(S^2) = \mathbb{Z}$, so even though the space is parallelizable there is a countable infinitude of non-homotopic classes of Lorentz structures.

Although every paracompact manifold admits a Riemannian metric, the same is not true of Lorentz metrics. However, if the manifold is non-compact then there will always be a non-zero vector field; hence, one will always have a Lorentz structure. For compact manifolds the obstruction to the existence of a global section of PT(M) is the Euler-Poincaré characteristic. Note that this is also the obstruction to the existence of a non-zero vector field, even though a line field is more general than a non-zero vector field.

We may also regard a Lorentz structure as a one-dimensional differential system that is defined by the chosen line field. By Frobenius, each such choice of line field defines a one-dimensional foliation of M by curves that are, *by definition*, timelike.

Yet another way of characterizing the line field that defines a Lorentz structure is derived from the fact that the projective cotangent bundle PT*(M) has a contact structure. A section of this bundle then becomes a Legendrian submanifold of a contact manifold. We shall return to this in the context of geodesic flows.

To summarize, we state:

**Theorem:**

*The following are equivalent to defining a Lorentz structure on a manifold* M:

1. *A symmetric non-degenerate second-rank covariant tensor field of hyperbolic normal type.*
2. *A global section of* PT(M).
3. *A one-dimensional dynamical system on* M.
4. *A Legendrian submanifold of the contact manifold* PT(M).
5. *A reduction of* GL(M) *to* O(3,1)(M).

**4. Proper-time foliations of Lorentz manifolds.** Suppose we have chosen a line field L on M as our means of defining a Lorentz structure. We return to the fact that L also represents a differential system, hence, a one-dimensional foliation on M, but with no canonical parameterization for any of the integral leaves; for the sake of reference, we shall call such a foliation a *proper time congruence.* Since there is nothing canonical about the parameterizations that one gives the leaves of this foliation, trying to find "transversal sections" of it, i.e., global simultaneity hypersurfaces, is more involved than simply looking at all points on all integral curves with the same value of curve parameter.

One way to approach the problem of defining transversal sections of the proper time congruence is to use the Lorentz metric to define an $n-1$-dimensional sub-bundle $\Sigma(M)$ of spacelike orthonormal complements to the line bundle, i.e., a Whitney sum splitting: $T(M) = \Sigma(M) \oplus L(M)$. Note that this also defines a reduction of O(1,$n-1$)(M) to O($n-1$)(M) by restricting oneself to all Lorentz frames that have a member in L(M). If this sub-bundle is integrable then its integral submanifolds define the simultaneity hypersurfaces. Nevertheless, the problem of integrability for this spacelike sub-bundle is not solved as simply as with the integration of the timelike line bundle. Note that just as the line field defines the time axis in the tangent space at every point, the orthogonal complement defines its $t = 0$ hyperplane, which also seems to have a distinctly Newtonian significance.

Indeed, if the foliation defined by $\Sigma(M)$ is to play the same role globally as the proper time of a fixed observer at a point $x$ we should expect the foliation to agree, in some way, with the proper time foliation of the tangent space to $x$ by the metric at that point. We see that locally the leaves of $\Sigma(M)$ look like the Newtonian foliation. We did show that this foliation was concordant to the proper time foliation of the open time cone. The real issue then becomes one of whether this local concordance can be extended to a global one.

Since a Lorentz metric defines a connection on T(M), or SO(3,1)(M), namely, the Levi-Civita connection, and it, in turn, defines an exponential map, which is a diffeomorphism from some neighborhood of the origin in any $T_x(M)$ onto a neighborhood of $x$, there should be at least a local image of the proper time foliation of $T_x(M)$. The usual issues pertaining to the exponential map, such as whether it is defined on all of $T_x(M)$ (i.e., geodesic completeness), whether it is injective (the existence of conjugate points), and whether it is surjective (geodesic connectedness) have corresponding issues pertaining to the proper time foliation of $T_x(M)$. In particular, a key issue is the extent to which the leaves of the local foliation at $x$ can be extended before the topology of M obstructs them. One should point out that the character of the exponential map for a Lorentz manifold is more complicated than it is for a Riemannian manifold since the Hopf-Rinow theorem does not hold for Lorentz manifolds, in general [**27, 28**].

In light of the foregoing, since a foliation that is obtained by integrating Σ(M) (assuming that bundle is integrable, of course) seems to be more closely related to the foliation of Minkowski space by its time coordinate − a distinctly Newtonian way of looking at time − than the proper time foliation, one is naturally suspicious of whether such a foliation even exists globally. Perhaps defining a local proper time foliation has something of the same general character as defining a local gauge for a field theory, and the extension to a global definition must be treated just as carefully. In particular, one might expect that the Godbillon-Vey class would play a crucial role, at least for spacetimes with non-vanishing $H^3(M, \mathbb{R})$, such as non-simply connected ones. (Such spacetimes are commonly thought to admit cosmic strings.)

**5. Time orientability.** One case in which the integration of the spacelike sub-bundle is simplified immensely is when the Lorentz manifold is assumed to be *time-oriented*. This means that there is a non-zero vector field **t** on M that spans the timelike line bundle L(M). Again, for non-compact manifolds this is no restriction, but for compact manifolds, the Euler-Poincaré characteristic $\chi(M)$ must vanish. However, this was already a condition for the existence of the Lorentz structure. Nevertheless, although the vanishing of $\chi(M)$ implies the existence of *some* non-zero vector field on M, it does not guarantee the existence of a non-zero section of L(M) in particular. Since a non-zero vector field will span a line bundle, hence, will define a section of PT(M), we are back to the previous question of whether all sections of PT(M) are homotopic, which they are not, in the general case.

Just as one can define a simply connected orientable covering manifold for a non-orientable manifold, can also define a simply connected time-orientable covering manifold of a non-time-orientable M. The fiber of any point $x \in$ M is then the pair of unit vectors **t** and −**t** at $x$ that span $L_x$(M). The covering map essentially represents the covering of the projective space in each tangent space by the unit sphere (relative to any auxiliary *Riemannian* metric that agrees with the Lorentz metric on L(M)) in each tangent space. A time orientation of M then becomes a section of the covering map; this also implies that any simply connected manifold is time oriented.

Once we have **t** to generate L(M) the proper-time congruence can be expressed as the orbits of the flow of **t**, either locally, or, if M is compact, globally. The question of finding a section of that flow is still not completely resolved, however, since a section of the proper time flow is still a codimension-one foliation of M by spacelike submanifolds.

Since **t** is everywhere non-zero, it can be normalized to have unit length everywhere. This has the effect of reducing the number of equivalent parameterizations of each integral curve to those that differ by a constant translation along the proper time axis, i.e., a choice of "zero proper time."

If we use the Lorentz metric to define the 1-form dual to **t** by $\theta = i_{\mathbf{t}} g$ (i.e., $\theta(\mathbf{v}) = g(\mathbf{t}, \mathbf{v})$) then we can define the complementary spacelike bundle Σ(M) by the annihilating subspaces to $\theta$, i.e., the subspaces of vectors **v** in each tangent space with the property that $\theta(\mathbf{v}) = 0$. Frobenius now takes the form of saying that Σ(M) is integrable into a framed foliation by spacelike leaves iff $\theta \wedge d\theta = 0$.

A simple way to satisfy this is to require that $d\theta = 0$. If one uses a hydrodynamical analogy for the proper time flow, in which $\theta$ represents either the *covelocity* (*momentum,* resp.) of the flow ([12]), then $d\theta$ is the *vorticity* (kinematical or dynamical, resp.) of the flow, and if the vorticity is zero the flow is said to be *irrotational* ([13]). In such a case, one is essentially ruling out closed timelike curves, i.e., proper time flow curves that are diffeomorphic to circles. However, one can still have integrable foliations defined by vortical flows, since the general case is that $d\theta = \alpha \wedge \theta$ for some 1-form $\alpha$.

For the irrotational case, we have two possibilities to consider: either $\theta$ is exact or not. If $\theta$ is exact then $\theta = dS$ for some smooth function S on M. In this case, we have our proper time foliation defined by level surfaces of S, up to a constant, which represents the ambiguity in the proper time origin. If $\theta$ is closed, but not exact, then no such S can exist, and the leaves of the proper time foliation are more involved. Clearly, the difference is meaningful only if $H^1(M)$ is not zero, such as when M is not simply connected, since otherwise all closed forms would be exact. Since this is true locally by the Poincaré lemma, we recover the fact that codimension-one foliations locally look like the level surfaces of smooth functions with no critical points. Note that in the event that M is not simply connected then, by Poincaré duality, we should have: $H^1(M, \mathbb{R}) \cong H^3(M, \mathbb{R})$. Since the Godbillon-Vey class lives in $H^3(M, \mathbb{R})$, this implies that there might be inequivalent foliations of non-simply connected spacetimes. In the simply connected case, the Hurewicz isomorphism theorem does not guarantee that $H^1(M, \mathbb{R})$ is trivial however, only that the first non-zero $\pi_k(M)$ is isomorphic to the corresponding $H^k(M, \mathbb{R})$. By Poincaré duality, if $k=3$ this could entail that $H^1(M, \mathbb{R})$ is non-zero, even if $\pi_1(M)=0$. Consequently, to have inequivalent foliations of spacetime, it is sufficient that the first non-vanishing homotopy group be $\pi_3(M)$. (Such a spacetime would admit "textures," but not "walls, strings, or monopoles.")

In looking at the problem of physically interpreting the transformation from one proper time foliation to another, one must be careful to distinguish two distinct situations: the transformation from one local chart of a *given* proper time foliation to another overlapping one, and the transformation from one proper time foliation to *another* such foliation; the latter case is closely related to the freedom to choose the line field that defines the Lorentz structure.

In the first situation, choose a line field L(M) and a corresponding Whitney sum decomposition of $T(M) = L(M) \oplus \Sigma(M)$, and assume that the spacelike sub-bundle $\Sigma(M)$ is integrable. A chart of the resulting codimension-one foliation of M is then an open subset $U \subset M$ and a differentiable submersion $\tau_U: U \to \mathbb{R}$, i.e., $D\tau_U|_x \neq 0$ for all $x \in U$, such that the foliation of U by the level surfaces of $\tau_U$ agrees with the foliation of M by integral leaves to $\Sigma(M)$. If $V \subset M$ is another open subset of M that overlaps U non-vacuously and $\tau_V: V \to \mathbb{R}$ is its associated foliation chart then by the definition of a foliation on M the overlap condition on U ∩ V says that for each $x \in$ U ∩ V there should be a diffeomorphism $\phi_{UV}(x)\colon \mathbb{R} \to \mathbb{R}$ such that for all $x \in$ U ∩ V one has $\tau_V(x) = \phi_{UV}(x) \circ \tau_U(x)$. Physically, this says that two local proper time charts can differ only by a reparametrization of the proper time line, but that they should still both have the same notion of simultaneity, i.e., the same integral leaves for the integration of $\Sigma(M)$.

---

[12]  The difference is geometrically meaningful only for the case of mass distributions with intrinsic angular momentum, for which the energy-momentum tensor $T_{ij}$ is not symmetric, or when $T_{ij}$ is degenerate. In either case, $T_{ij}$ would not define a metric tensor field.

[13]  One can also think of $\delta\theta$ − the codifferential of $\theta$ − as the *compressibility* of the flow.

Now consider the case of two proper time foliations of M, which we assume are defined by two distinct choices of line field $L_1(M)$ and $L_2(M)$, with corresponding spacelike sub-bundles $\Sigma_1(M)$ and $\Sigma_2(M)$. At every point $x \in$ M choose a (non-unique) Lorentz transformation that takes $L_1(x)$ to $L_2(x)$ – hence, by orthogonality, $\Sigma_1(x)$ to $\Sigma_2(x)$ – in such a way that one can smoothly piece together all of the individual Lorentz transformations into a (non-unique) vertical automorphism $\mathscr{V}$ of SO(3,1)(M), which essentially amounts to a "gauge transformation" as far as a choice of rest frame and simultaneity is concerned. It is not hard to see that the question of how the leaves of one foliation relate to those of the other is topologically deep; for one thing, one has to establish the sense in which the one foliation is "equivalent" to the other. In addition to the aforementioned integrable homotopy, concordance, and foliated cobordism, one could also simply look for a diffeomorphism of M to itself that takes any leaf of one foliation to a leaf of the other. Physically, the fact that a different choice of rest frame generally implies a "different" foliation is consistent with the relativistic notion that it should also define a different conception of the simultaneity of events.

The aforementioned theorem of Geroch [5] that implies that spacetime must be cylindrical makes critical use of time-orientability. What it says, specifically, is that if the disjoint union $\Sigma_1 \vee \Sigma_2$ of two closed (i.e., compact, without boundary) 3-manifolds bounds a compact time-oriented Lorentz 4-manifold M (i.e., $\Sigma_1$ and $\Sigma_2$ are *Lorentz cobordant)* that admits no closed timelike curves, then $\Sigma_1$ and $\Sigma_2$ are diffeomorphic, in which case, M = $\Sigma_1 \times [0,1]$. The method of proof is enlightening because it basically amounts to regarding the timelike vector field **t** that time-orients M as a gradient flow for a Morse function $\tau$, which is essentially a global proper time function. The fact that **t** has no zeroes on M implies that its flow has no fixed points in M. Compactness implies that an integral curve that starts on $\Sigma_1$ will end on $\Sigma_2$, and the flow will take $\Sigma_1$ diffeomorphically to $\Sigma_2$. If $\tau$ varies from 0 to 1 along the integral curves of **t** then the level sets of $\tau$ will define a cylindrical foliation of M. A corollary to this theorem is that closed time-orientable four-dimensional Lorentz manifolds must have closed timelike curves, a result found earlier by Bass and Witten, as well as Kronheimer and Penrose [29].

Another way of arriving at Geroch's theorem is to note that it is equivalent to a special case of Painlevé's theorem [7] that if M is a compact manifold with a framed codimension-one foliation, then if any one of its leaves is compact and simply connected, then all the leaves are diffeomorphic to it. (This makes the foliation a fibration over $S^1$.) Apparently, the key idea to appreciate is that time-orientability is a stronger statement than it sounds like when viewed by the eyes of topology, since the flow of a non-zero vector field on a compact manifold M is a global action of $\mathbb{R}$ on M by diffeomorphisms, i.e., a dynamical system. If one is to expect the appearance of non-diffeomorphic leaves and non-trivial Lorentz cobordisms, then apparently one has to expect some weakening of the assumptions, such as compactness or the non-existence of closed timelike curves.

A subtle point to consider regarding the construction of the vector field **t**, at least in the eyes of physics, is the issue of whether that vector field is assumed to originate from purely mathematical considerations or from more physically intuitive ones. Generally, physics tends to regard such a vector field as a purely mathematical device that one introduces without any physical considerations, but when one looks at the relationship between **t** and the concept of a rest frame for an extended observer, as well as the aforementioned hydrodynamical analogy for the character of the proper time flow, it becomes more plausible that the vector field **t** can *only* be physically defined by the motion of some physical entity.

This immediately brings the question of the support of **t** to the foreground, which, in turn, relates to the question of whether spacetime need be globally time-orientable in fact. Upon further contemplation, one sees that proper time is defined only where there the motion of some *physical* entity is defined. Of course, the concept of a "physical entity" can include things whose motion has an astronomical support, such as the light waves that are emitted from stars, or perhaps even the gravitational waves that have been theorized. However, in the eyes of topology, the very existence of even *one* point in all of spacetime at which such motion could not be defined – i.e., a spacetime singularity – would be a serious matter to accommodate.

The ultimate issue to consider is the issue of whether the entire spacetime manifold can indeed be obtained as the Cauchy development of some maximal Cauchy hypersurface $\Sigma$ for the integration of Einstein's equations as equations for the proper time evolution of a spacelike metric on $\Sigma$, or if the existence of singularities might obstruct the existence of a *maximal* Cauchy hypersurface. Certainly, if any of the more exotic topological methods in mathematical physics, such as cobordism and the theory of foliations, are to be absolutely unavoidable, one must be able to justify that the spacetime manifold is not merely cylindrical, as has been tacitly assumed in most of the work done in relativity that is concerned with the problem of the proper time evolution of the spacetime metric, such as the Hamiltonian formulation of general relativity [**6**].

Note that, in light of the remarks made earlier in the context of codimension-one foliations that are defined by smooth functions on closed manifolds, if a proper-time function is globally defined on a closed (perhaps compactified) spacetime manifold then it must have critical points. Hence, the appropriate foliation to consider would be the singular foliation or Haefliger structure. This would also suggest that proper-time evolution represents a "topology-changing process" as far as the simultaneity submanifolds are concerned.

We shall now look at how the problem of transversal sections of flows appears in various specific motions on a spacetime manifold. The examples we shall consider are the geodesic flow of the Lorentz metric, the Cauchy problem in its general sense, and its application to the various physical wave motion, such as electromagnetic and gravitational waves.

**6. Transversal sections of the geodesic flow of a Lorentz manifold.** As mentioned above, one place in which proper time foliations play an important role is in the question of whether a given flow admits a transversal section. This is equivalent to the issue of whether there is a universal rest frame for all of the integral curves of the flow. In particular, since the motion of uncharged matter in gravitational fields is by geodesics, we shall discuss that particular flow, although the issue of the integrability of the normal bundle to the dynamical system is also quite important in non-relativistic hydrodynamics, as well. First, we discuss the formulation of the geodesic flow as a Hamiltonian flow.

The cotangent bundle to any manifold – spacetime, in particular – admits a canonical symplectic structure. It is defined by starting with the canonical 1-form $\theta$ on T*(M), which is, in turn, defined by $\theta_p(\mathbf{v}) = p(\pi_* \mathbf{v})$ where $p$ is a cotangent vector (i.e., 1-form) on M and $\pi$ is the projection of the cotangent bundle onto M [*cf.* **30-35**].

$\theta$ defines a closed non-degenerate 2-form by way of $\Omega = d\theta$, i.e., a *symplectic structure* on T*(M). The nondegeneracy allows one to define a symplectic duality between tangent vectors and cotangent vectors by means of the linear isomorphism that takes a tangent vector $\mathbf{v}$ to the linear functional, $\varpi$, defined by $\varpi(\mathbf{w}) = \Omega(\mathbf{v}, \mathbf{w})$, i.e., $\varpi = i_\mathbf{v}\Omega$. In particular, the inverse of this fiberwise isomorphism takes any 1-form to a vectorfield, which can generate a dynamical system on M. In the case where the 1-form is exact, i.e., of the form $d$H for some 0-form H called a *Hamiltonian function,* the symplectic dual of $d$H, which we will notate by $X_H$, is a *globally Hamiltonian vectorfield.* Its flow is then called a Hamiltonian flow ([14]). We can also say that $X_H = \nabla_\Omega H$ is the *symplectic gradient* of H.

To make contact with the classical mechanical notations for Hamiltonian mechanics one needs to define a coordinate system on an open subset U of M, whose coordinate functions will be notated by $q^i$. This defines a local coframe by the 1-forms $dx^i$. Relative to this coframe, any 1-form $p$ on U is represented by $p = p_i dx^i$. Hence, a local trivialization of T*(U) is defined by $(x^i, p_i)$. Somewhat confusingly, the 1-form $\theta$ also takes the same form as $p$, $\theta = p_i dx^i$, even though $p$ is 1-form on M and $\theta$ is a *semi-basic* 1-form on T*(M); this means that its components are functions on T*(M), not functions on M. Moreover, when $\theta$ is pulled down to M by way of $p$, one gets $p: p^*(\theta) = p$. This makes $\Omega = dp_i {\wedge} dx^i$. Since we have the local expression:

$$d\mathrm{H} = \frac{\partial \mathrm{H}}{\partial x^i} dx^i + \frac{\partial \mathrm{H}}{\partial p_i} dp_i$$

for any 0-form H the Hamiltonian vectorfield obtained by $\Omega$ takes the local form:

$$X_H = \nabla_\Omega \mathrm{H} = \frac{\partial \mathrm{H}}{\partial p_i} \frac{\partial}{\partial x^i} - \frac{\partial \mathrm{H}}{\partial x^i} \frac{\partial}{\partial p_i}$$

Its integral curves $\gamma$ are defined by integrating Hamilton's equations, either in the general form $\dot{\gamma} = X_H$ or in the local form:

$$\frac{dq^i}{d\tau} = \frac{\partial \mathrm{H}}{\partial p_i}$$

$$\frac{dp_i}{d\tau} = -\frac{\partial \mathrm{H}}{\partial x^i} \, .$$

---

[14] When the 1-form is closed, but not exact, one speaks of a *locally Hamiltonian* vector field.

(Here, we use $\tau$ for a curve parameter to emphasize that the proper time curve parameter is distinct from the time coordinate.)

Since any smooth function will do for H, it is reasonable to consider using functions on T*(M) that are defined by the Lorentz metric. In particular, Q defines such a function on T*(M) if one defines the Lorentz metric $g$ on T*(M) instead of T(M). (Since $g$ defines a metric isomorphism of tangent and cotangent spaces, this is no problem.) Locally the Lorentz structure on T*(M) looks like:

$$g = g^{ij}\mathbf{e}_i \otimes \mathbf{e}_j,$$

relative to a local frame field $\mathbf{e}_i$. Hence, we define our Hamiltonian H using the associated quadratic form: $H(p) = \frac{1}{2}g(p, p) = \frac{1}{2}g^{ij}p_ip_j$. This makes the Hamiltonian vector field look like:

$$X_Q = g^{ij}p_j\frac{\partial}{\partial x^i} - \frac{1}{2}\frac{\partial g^{ij}}{\partial x^k}p_ip_j\frac{\partial}{\partial p_k}.$$

The corresponding Hamilton equations become:

$$\frac{dq^i}{d\tau} = g^{ij}p_j$$

$$\frac{dp_i}{d\tau} + \frac{1}{2}\frac{\partial g^{ij}}{\partial x^k}p_ip_j = 0.$$

The first equation amounts to a definition in terms of metric duality. The second equation looks suspiciously similar to an equation of parallel translation for $p$ for some sort of metric connection, so we pursue this further.

Regard the metric $g$ on T*(M) as an SO(1, 3)-equivariant map from the bundle GL(M) of oriented Lorentz coframes on spacetime to the space of Lorentz metrics on $\mathbb{R}^4$. This space is diffeomorphic to the homogenous space GL$^+$(4)/SO(1,3), whose elements are best represented by symmetric non-degenerate 4×4 real matrices. All this says is that $g$ associates a matrix of components $g^{ij}$ to every frame $\mathbf{e}_i(x)$ in T(M). For each element $\mathbf{e}_i \in$ GL(M), the metric $g$ on T*(M) looks like $g = g^{ij}\mathbf{e}_i \otimes \mathbf{e}_j$. To deal with the problem of differentiating frames, we introduce a connection $\omega_i^j$ on GL(M):

$$D\mathbf{e}_i = -\omega_i^j\mathbf{e}_j = \Gamma_{ik}^j\theta^k \otimes \mathbf{e}_j.$$

One thing must be pointed out before we go further: if $\omega_j^i$ is a connection that makes:

$$\frac{1}{2}\frac{\partial g^{ij}}{\partial x^k}p_ip_j = \Gamma_{kj}^i\dot{x}^jp_i = \Gamma_{km}^ig^{mj}p_ip_j,$$

then it is not unique. However, the Hamiltonian vector field $X_Q$ is defined only in terms of the metric and the canonical 1-form on T*(M); hence, $X_Q$ *is* unique. Indeed, there is an equivalence class of connections on the Lorentz frame bundle that produce the same $X_Q$. Let us pick a metric connection:

$$0 = dg\ = dg^{ij}\mathbf{e}_i \otimes \mathbf{e}_j + g^{ij}d\mathbf{e}_i \otimes \mathbf{e}_j + g^{ij}\mathbf{e}_i \otimes d\mathbf{e}_j$$
$$= (dg^{ij} - g^{ik}\omega_k^j - g^{ik}\omega_k^j)\ \mathbf{e}_i \otimes \mathbf{e}_j.$$

This makes:

$$dg^{ij} = g^{ik}\omega_k^j + g^{jk}\omega_k^i$$

so:

$$dg^{ij}p_ip_j = g^{ik}\omega_k^j\,p_ip_j + g^{jk}\omega_k^i\,p_ip_j = 2\,\omega_j^i\dot{x}^j\,p_i,$$

hence,

$$\tfrac{1}{2}\frac{\partial g^{ij}}{\partial x^k}\,p_i\,p_j = \Gamma_{kj}^i\dot{x}^j\,p_i,$$

and the second Hamilton equation takes the form:

$$\dot{x}^j\left(\frac{\partial p_k}{\partial x^j} - \Gamma_{kj}^i\,p_j\right) = 0.$$

The only thing stopping us from calling this the equation of parallel translation for this connection is that the lower indices on $\Gamma_{kj}^i$ are "backwards". To remedy this, we need to choose the torsion 2-form for the connection. The usual choice is zero, i.e., the Levi-Civita connection, with $\Gamma_{jk}^i = \Gamma_{kj}^i$, in which case, the second Hamilton equation does indeed give the equation of parallel translation for $p$ along an integral curve of $\dot{x}$. However, one is metric dual to the other, so such a curve would be a geodesic of the metric $g$ on M.

All of the foregoing basically justifies our declaring $X_Q$ to be the *geodesic spray* of $g$ on T*(M), and its flow to be the *geodesic foliation* of T*(M) that is defined by $g$, although customarily one uses those terms for the corresponding constructions on T(M); our definition on T*(M) simply takes advantage of the natural symplectic structure on T*(M), since a symplectic structure seems more contrived when defined on T(M).

To turn this foliation on T*(M) into a *geodesic flow* [36] on M it is not enough to merely project the flow from T*(M) onto M since geodesics though a given point $x \in$ M that have distinct values of $p(x)$ will still project onto the same $x$, i.e., the resulting curves would intersect, which would violate the definition of a foliation([15]). Moreover, a section $p: M \rightarrow T*(M)$ with the property that if $\mathbf{v}$ is the vectorfield that is metric-dual to $p$, so $p = i_{\mathbf{v}}g$, then the image of every integral curve of $\mathbf{v}$ under $p$ must be an integral curve of $X_Q$. (This simply says that the integral curves of $\mathbf{v}$ are obtained by integrating $X_Q$ twice.) As a consequence, by differentiation we also have $Dp|_x(\mathbf{v}) = X_Q(p(x))$, i.e., $p*(\mathbf{v}) = X_Q$ when restricted to the image of $p$. We call such a section a *geodesic section* of T*(M) even though that term has a different definition in variational calculus, and proceed to justify the use of the term.

In its usual variational sense [37-40], when a variational field problem is defined by a semi-basic $n$-form $\omega$, on a fiber bundle E $\rightarrow$ M over an $n$-dimensional manifold M, a section $\phi$ of E is a *geodesic section* iff $\phi*\omega$ is closed, i.e., $\phi*d\omega = 0$. In the present case of one-dimensional foliations, if the semi-basic 1-form is $\theta$, then for a section $p$ of T*(M) to be geodesic we must have: $d(p*\theta) = dp = 0$, since the definition of $\theta$ makes $p^*\theta = p$. Hence, $p$ is a geodesic section of T*(M) iff it is closed. This tends to suggest that flows with non-zero vorticity cannot be geodesic in this sense.

---

[15] Nevertheless, these projective singularities can be interesting in their own right, cf., Sardanashvily [9-11].

Note that in the previous paragraph we made no reference to the Hamiltonian Q, as we did in the paragraph before that, but only to the canonical 1-form $\theta$ on T*(M). To reconcile the two ways of characterizing geodesic sections of T*(M), we have to include Q in the definition of $\omega$. Since $\omega$ is the 1-form which, when pulled down by $p$ and integrated along a path, gives the action along that path, and, for a velocity vector field $v(\tau)$ that is tangent to the path, $\omega$ amounts to the Lagrangian 1-form $\mathcal{L}(v(\tau))d\tau$ the proper way to define $\omega$ (at least locally) is by the Poincaré-Cartan [**30**] form:

$$\omega = p_i dx^i - Q d\tau.$$

Here, the index $i$ is summed over the spacelike indices, and $d\tau$ is the 1-form that is metric dual to the timelike vector field that defines the foliation. From this, we get:

$$d\omega = dp_i {}^\wedge dx^i - dQ {}^\wedge d\tau.$$

In order to pull this expression down by $p$ (along a curve), we express $dp_i$ and $dQ$ in terms of $dx^i$ and $d\tau$.

$$dp_i = \frac{dp_i}{d\tau} d\tau, \qquad dQ = \frac{\partial Q}{\partial x^i} dx^i + \frac{\partial Q}{\partial x^i} \frac{dp_i}{d\tau} d\tau.$$

Substituting in $d\omega$ gives the pull-back of $d\omega$ by $p$ as:

$$p^* d\omega = \left( \frac{dp_i}{d\tau} + \frac{\partial Q}{\partial x^i} \right) d\tau {}^\wedge dx^i.$$

Setting this to zero gives half of Hamilton's equations. Combining this with the definition of $p$ as being the metric dual to a velocity vector field:

$$\frac{dx^i}{d\tau} = g^{ij} p_j,$$

gives the equivalence of the two ways of defining geodesic fields.

In the event that $p$ is a geodesic section of T*(M), the image of M under the section $p$ would also have the property that $\Omega = d\theta = 0$ when restricted to $p$(M). Since $p$(M) is of dimension $n$ and the rank of $\Omega$ restricted to it is $n$, this makes $p$(M) a *Lagrangian submanifold* of T*(M). Another example of a Lagrangian submanifold of T*(M) is its zero section, which is essentially the special case where $p = 0$. However, since $p$ was supposed to be dual to a non-zero vector field, this cannot be the case. One then looks at other Lagrangian submanifolds as diffeomorphisms of the zero section, and the projective singularities previously alluded to are what one encounters when such a diffeomorphism does not cover the identity under the bundle projection, i.e., preserve the fibers of T*(M).

So far we have restricted our class of covelocity (or momentum) 1-forms $p$ to those that are closed. If we further suppose that p is exact $p = d$S the definition of a level surface for our Hamiltonian becomes a first order PDE whose solution is S, namely, the *relativistic Hamilton-Jacobi equation*:

$$\mathrm{H}_0^2 = \tfrac{1}{2}\, g(d\mathrm{S}, d\mathrm{S}) = \tfrac{1}{2}\, g^{ij} \frac{\partial \mathrm{S}}{\partial x^i} \frac{\partial \mathrm{S}}{\partial x^j},$$

which is also called the *eikonal equation* in optical terminology. The levels surfaces of S define *simultaneity hypersurfaces.* Their gradient vectorfield is **t**, whose integral curves are geodesics transversal to the hypersurfaces in a manner analogous to the relationship between light rays and light waves. The geodesics define a foliation that is transverse to and complementary to the one defined by S, and is referred to as the *Caratheodory complete figure* [37- 39] for the Hamilton-Jacobi problem. Since any two such geodesics intersecting two hypersurfaces will have the same proper time length, one calls the hypersurfaces *geodesically equidistant.*

The fact that **t** is non-zero implies that all of the level surfaces of S are diffeomorphic, so clearly this case is relevant only to the consideration of sectionable proper-time flows, i.e., cylindrical spacetimes. Conversely, if spacetime is cylindrical $\mathrm{M} = \Sigma \times \mathbb{R}$, and we define S to be the projection on $\mathbb{R}$, then $d$S satisfies the eikonal equation. This suggests that the only way in which non-cylindrical proper-time foliations of spacetime would become interesting would be in the case where $p$ is not exact, i.e., **t** is not a gradient vector field. For example, this would be the case if $p$ were not closed (**t** were not irrotational), although that would involve the possibility of closed timelike geodesics, which are generally symptomatic of a serious pathology in causality.

For the aforementioned proper time foliation of Minkowski space, since we are moving in a vector space, the connection that gives us the geodesic spray is zero and the Hamilton-Jacobi formulation of this scenario is rather uninteresting. The solution to the Hamilton-Jacobi equation $(\pm \tau^2 = g(d\mathrm{S}, d\mathrm{S}))$ is simply $\mathrm{S} = \tfrac{1}{2}\, \mathrm{Q}$, and the characteristic equations $(\dfrac{dx}{d\tau} = p,\ \dfrac{dp}{d\tau} = 0)$ define unaccelerated motion, so the pair of complementary foliations that are defined by the Lorentz orbits and their orthogonal trajectories represent the Caratheodory complete figure.

Although we have been referring to global sections of T*(M), if we consider the physical problems in which geodesic flows are appropriate then we can see that we should distinguish between local and global sections more carefully, especially because there might be topological obstructions to the extension of local sections to global ones. For the case of relativistic hydrodynamics, if $p$ represents the energy-momentum 1-form for a fluid then it is only necessary for the geodesic section $p$ to have compact spacelike support. The spacetime support of $p$ is then referred to as a *flow tube* for the motion. In such a situation it is irrelevant whether a global extension of the support exists, since one does not expect mass distributions to fill the entirety of spacetime. The situation in which the existence of a global geodesic section has any physical sense is then concerned with the issue of time orientability. If one requires that the vector field **t** that orients the line bundle that defines the Lorentz structure be a geodesic vector field, so $\nabla_t \mathbf{t} = 0$, then its metric dual $i_t g$ will be a global geodesic section of T*(M). However, since the zero section of T*(M) is itself a geodesic section, in the non-time-orientable case a singular geodesic **t** will still give a geodesic section of T*(M).

**7. Discussion.** Considering the depth, breadth, and complexity of both the theories of foliations and spacetime structure, it is, of course, impossible to completely span all of the issues in one article. The primary purpose of the foregoing was to examine the subtleties that are associated with trying to extend the proper time foliation of Minkowski space to a corresponding foliation of a Lorentz manifold.

Since the essence of the physical justification for the methods of foliations lies in the nature of simultaneity and proper time, it stands to reason that the deeper matters of Lorentz causality will also be relevant. In particular the role of spacetime singularities is important. In this case, we mean not only the singularities of the metric itself, i.e., points where the metric might be degenerate, in which case, the Lorentz structure itself would break down, but also points where time-orientability is violated, and points where the geodesic flow has projective singularities; respectable progress in this direction has been made by Sardanashvily, as noted.

Insofar as foliations generalize dynamical systems, a large part of the research available in the mathematical literature is concerned with leaf dynamics, that is, the way that the topology of the leaves in the codimension-1 case varies with "time" (which is not necessarily a parameter, of course). This clearly implies subtle ramifications in initial value problems, such as gravity and electromagnetism. Some progress has been made in the gravitational context as regards the issues of constant-curvature foliations, maximal hypersurfaces, the Raychaudhuri equation [**41-44**]. The methods of foliations also provide another way of modeling the intriguing notion of "topology-changing processes."

One of the points that undoubtedly will provide deeper insights into the way that spacetime structure affects the structure of physical models is the topic we ultimately alluded to, of the role of foliations as generalized gauge structures, and how proper time reparameterizations relate to the current algebras of particle physics. Perhaps that path will lead further in the direction of defining the presence of matter in a spacetime, as well as its structure, as a *consequence* of deeper considerations in the geometrical and topological structure of spacetime.

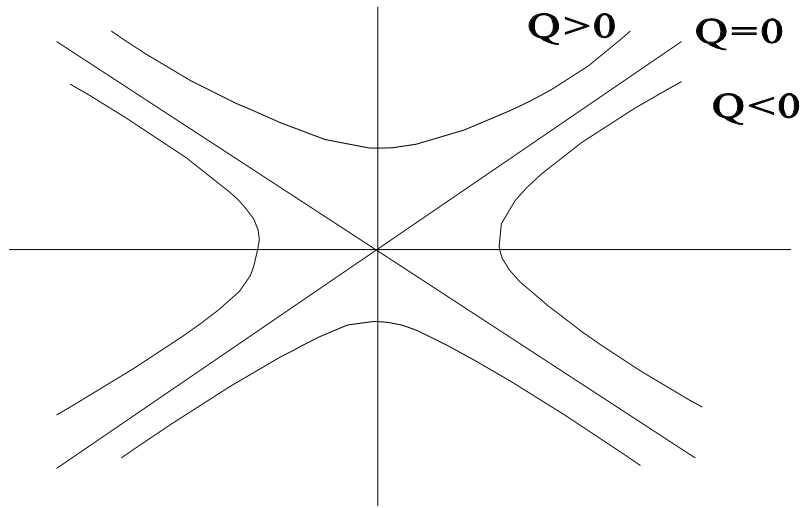

Fig. 1.  Proper-time foliation of Minkowski space ($Q = \tau^2$).

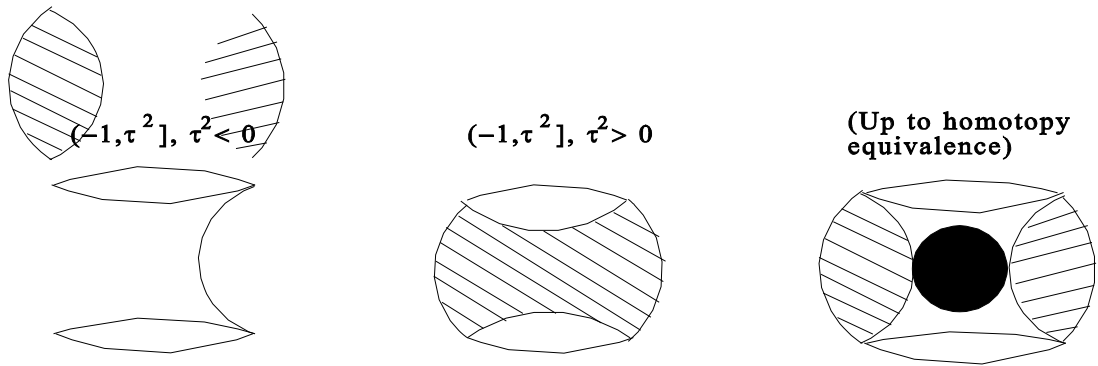

Fig. 2. Attaching an n-1-cell as $\tau^2$ passes the critical value of 0, $\eta = (+ - - \ldots -)$.

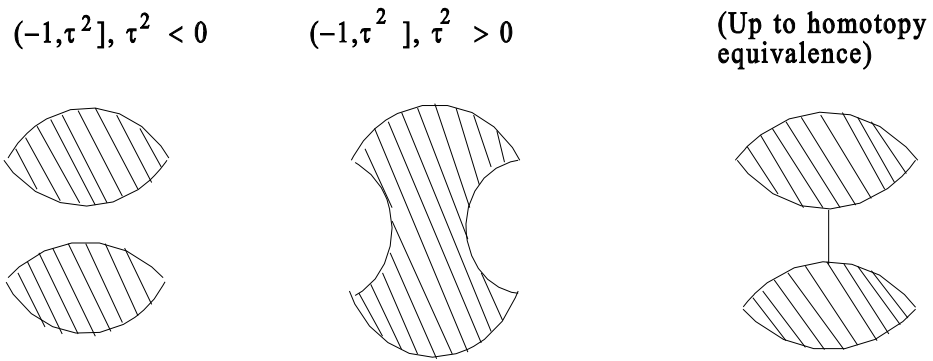

Fig. 3. Attaching a 1-cell as $\tau^2$ passes the critical value of 0, $\eta = (- + + \ldots +)$.